\documentclass[aps,prb,twocolumn,floats,english]{revtex4}
\usepackage{graphicx}
\usepackage{amssymb}

\makeatletter

\providecommand{\tabularnewline}{\\}

\usepackage{babel}
\makeatother
\begin{document}

\title{Connecting systems with short and long ranged interactions: local molecular field
theory for ionic fluids}

\author{Yng-gwei Chen$^{1,2}$, Charanbir Kaur$^{1}$, and John D. Weeks$^{1,3}$}

\affiliation{$^{1}$Institute for Physical Science and Technology and\\
 $^{2}$Department of Physics and\\
 $^{3}$Department of Chemistry and Biochemistry\\
 University of Maryland, College Park, Maryland}

\date{\today}

\begin{abstract}
Structural and thermodynamic properties of ionic fluids are related to
those of a simpler ``mimic'' system with short ranged intermolecular
interactions in a spatially varying effective field by use of Local
Molecular Field (LMF) theory, already successfully applied to nonuniform
simple fluids. By consistently using the LMF approximation to describe
only the slowly varying part of the Coulomb interaction, which we view as
arising from a rigid Gaussian charge distribution with an appropriately
chosen width $\sigma$, exceptionally accurate results can be found. In
this paper we study a uniform system of charged hard spheres in a
uniform neutralizing background, where these ideas can be presented in
their simplest form. At low densities the LMF theory reduces to a
generalized version of the Poisson-Boltzmann approximation, but the
predicted structure factor satisfies the exact Stillinger-Lovett moment
conditions, and with optimal choice of $\sigma$ the lowest order
approximation remains accurate for much stronger couplings. At high
density and strong couplings the pair correlation function in the
uniform mimic system with short ranged interactions is very similar
to that of the full ionic system. A simple analytic formula can then
describe the difference in internal energy between the ionic system and
the associated mimic system.
\end{abstract}

\maketitle
\section{Introduction}

In this paper we describe a new theory for the structure
and thermodynamics of ionic fluids based on a generalization of the
local molecular field (LMF) theory we have successfully applied to
nonuniform simple fluids.\cite{katsov.k.weeks.jd:incorporating,
katsov.k.weeks.jd:on,weeks.jd.katsov.k.ea:roles,weeks.jd.selinger.rlb.ea:self-consistent,weeks.jd.vollmayr.k.ea:intermolecular,weeks.jd:connecting}
A basic step in the LMF theory is the replacement of longer ranged
and slowly varying parts of the intermolecular interactions by an
appropriately chosen effective single particle potential. The structure
and thermodynamics of the resulting reference system
with shorter ranged intermolecular interactions in the presence of the
effective field is then related to that of the original system of
interest. 

This strategy seems particularly appropriate for ionic systems since
at long distances the Coulomb interaction is weak and very slowly
varying, and systems with short ranged interactions are significantly
easier to treat by theory or simulations. However Coulomb interactions
can be very strong and rapidly varying at short distances. A key question
we address is how to divide the Coulomb interaction into ``short''
and ``long'' ranged parts so that the LMF theory can give accurate
results. Its answer allows for the first \emph{controlled} use of
the LMF approximation, and we find exceptionally accurate results,
even better than those found earlier for fluids with short ranged
interactions.

Although the most physically interesting applications of these ideas
are probably to nonuniform mixtures of size and charge asymmetric
ions, in this initial discussion we consider a uniform \emph{one component
charged hard sphere system} (OCCHS) where almost all the ideas in
the LMF theory can be seen in their simplest form. \cite{hansenmac}
The OCCHS is made
up of (say positively) charged ``ions'' comprised of hard spheres
with a diameter $d\geq0$ with embedded positive point charges in
the presence of a uniform neutralizing background. The only nontrivial
correlations are between the positive ions and for most purposes we
can think of this as a one component system with very long ranged
repulsive interactions. A special case is the \emph{one component
plasma} (OCP) where there is no hard core ($d=0$). Nothing in the
theory makes essential use of the simplifications in the OCCHS. Generalizations
to nonuniform and asymmetric models are straightforward in most cases,
and equally good results have been found. 

\section{Local Molecular Field Equation}

\label{sec:lmf}

\subsection{Nonuniform systems}

We first discuss the qualitative ideas leading to the LMF equation.
These will be further developed and made more precise in our
discussion of the OCCHS. The simplest application of LMF theory relates
the structure and thermodynamics of a nonuniform system of interest
with a spherically symmetric pair potential $w(r)$ in an external
field whose value at any point $\textbf{r}_{1}$ is
$\phi({\textbf{r}}_{1})$ to those of a \emph{reference system} with
a shorter ranged pair interaction $u_{0}(r)$ in a renormalized effective field
$\phi_{R}({\textbf{r}}_1)$.  $\phi_{R}$ is supposed to be
chosen to take account of the averaged effects of
the \emph{perturbation potential} $u_1(r)$, where
\begin{equation}
w(r)=u_{0}(r)+u_{1}(r).\label{eq:wdiv}
\end{equation}

This separation of the intermolecular interaction $w$ into two parts can
be done in an infinite number of ways, and any choice of $u_1$ defines
a possible associated reference system. However
the averaging procedure leading to the simple LMF theory can be expected
to give very accurate results only for certain
properly chosen \emph{slowly varying} $u_1$.

Figure 1 gives examples of separations of the repulsive Coulomb
potential we will use in this paper, parameterized by an important
length scale $\sigma$. As explained in detail in Section \ref{sec:lmfocchs},
when $\sigma$ is chosen larger than some state-dependent minimum size
$\sigma_{\min}$, the Coulomb perturbation $u_1$ is sufficiently slowly
varying that the LMF theory can give very accurate results.
This is the crucial step in developing a simple and accurate theory
for ionic systems.

We will refer to the resulting special reference
systems with properly chosen $\sigma$ as ``mimic systems.'' 
In realistic models of ionic solutions
there are always strong short ranged repulsive core interactions that must be
dealt with in any quantitative theory or simulation. The mimic system simply
treats the short ranged rapidly varying part of the Coulomb potential as an
additional core-like contribution that generates a modified core interaction.
As we will see, many
properties of the full long ranged system can be very accurately described using
those of the short ranged mimic system.

\begin{figure}
\includegraphics[scale=0.32]{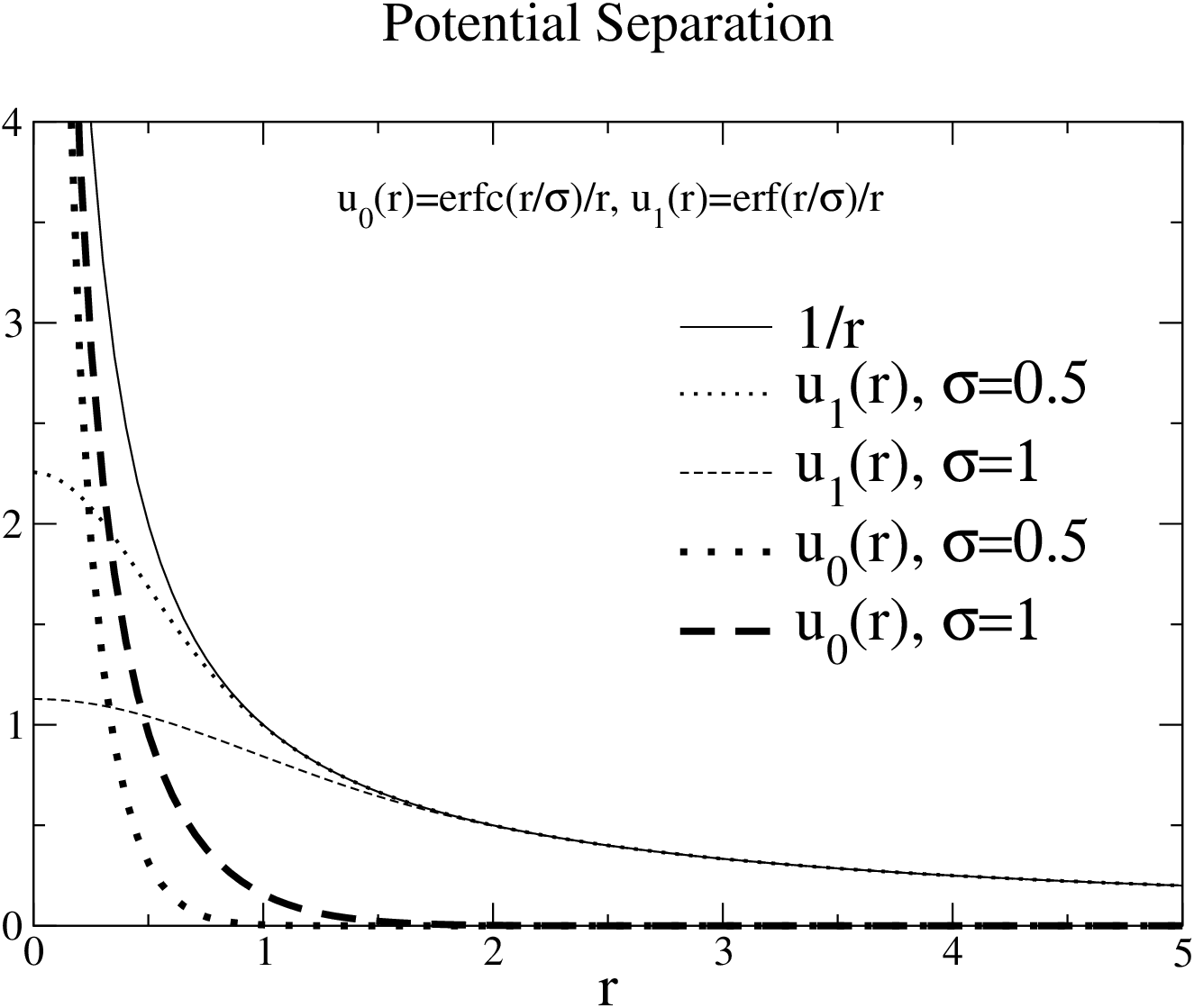}
\caption{\label{cap:potsept} The separation of the $1/r$ potential into a
short ranged piece $u_{0}(r)=\textrm{erfc}(r/\sigma )/r$ and a
long ranged piece $u_{1}(r)=\textrm{erf}(r/\sigma )/r$. A bigger
$\sigma $ corresponds to a longer ranged mimic system $u_{0}(r)$, and a
more slowly varying $u_{1}(r)$. Here two relevant $\sigma $ values are
shown for comparison.}
\end{figure}

For simple fluids with short ranged interactions the LMF approach has
proved most useful when $w$ can be divided into a slowly varying
perturbation $u_{1}(r)$ describing the relatively weak and longer ranged
attractive interactions and a short ranged rapidly varying core
potential $u_{0}$, which accounts for the local excluded volume
correlations of the particles. \cite{weeks.jd.chandler.d.ea:perturbation}
A separation with these qualitative features suffices to motivate the
derivation of the basic LMF equation that follows.

For any given $\phi({\textbf{r}})$ an associated
$\phi_{R}({\textbf{r}})$ could always found in principle so that the
nonuniform singlet density $\rho_{0}({\textbf{r}};[\phi_{R}])$ in the
reference system (denoted by the subscript $0$) equals that in the full
system $\rho({\textbf{r}};[\phi])$. Of course the latter is not known in
advance and its determination is one of the main goals of the theory.
However, if a $u_{1}(r)$ can be chosen to be
slowly varying over the range of excluded
volume correlations induced by the short ranged potential $u_{0}(r)$,
then we can make some physically motivated approximations to derive a
self-consistent equation to determine the associated $\phi_{R}.$

At high densities we expect that short ranged correlations in both
systems are controlled by packing effects from the identical repulsive
cores, and it seems plausible that $\phi_{R}$ can be chosen so that
both the singlet densities and the conditional singlet densities in
the reference and full systems resemble each other. That is, when
\begin{equation}
\rho_{0}({\textbf{r}};[\phi_{R}])=\rho({\textbf{r}};[\phi]),\label{eq:rho0rho}
\end{equation}
we also expect that
\begin{equation}
\rho_{0}({\textbf{r}}_{1}|{\textbf{r}}_{2};[\phi_{R}])
\simeq\rho({\textbf{r}}_{1}|{\textbf{r}}_{2};[\phi])\label{eq:crho0crho}
\end{equation}
holds to a good approximation. Here
$\rho_{0}({\textbf{r}}_{1}|{\textbf{r}}_{2};[\phi_{R}])$
is the (conditional) density at ${\textbf{r}}_{1}$ given that a particle
is fixed at ${\textbf{r}}_{2},$ directly related to the nonuniform
pair correlation function. With this assumption we can derive an equation
for $\phi_{R}$ that also turns out to give exact results at very
low densities, where pair correlations are not important. 

As discussed previously,\cite{weeks.jd.selinger.rlb.ea:self-consistent}
by subtracting the balance of forces as described
by the exact Yvon-Born-Green hierarchy\cite{hansenmac}
for the full and reference systems we find
a relation between the associated forces
\begin{equation}
-\nabla_{1}[\phi_{R}({\textbf{r}}_{1})-\phi({\textbf{r}}_{1})]=
-\int d{\textbf{r}}_{2}\rho_{0}({\textbf{r}}_{2}|{\textbf{r}}_{1};[\phi_{R}])
\nabla_{1}u_{1}(r_{12}).\label{eq:wsb}
\end{equation}
Moreover, if $u_{1}(r_{12})$ is very slowly varying over the range
of short ranged pair correlations, then $\nabla_{1}u_{1}(r_{12})$
essentially vanishes in the range of integration where
$\rho_{0}({\textbf{r}}_{2}|{\textbf{r}}_{1};[\phi_{R}])$
differs significantly from $\rho_{0}({\textbf{r}}_{2};[\phi_{R}])$
in eq \ref{eq:wsb}. Then we can replace the former by the latter,
and take the gradient outside the integral and integrate eq
\ref{eq:wsb}.

In making this replacement we have ignored correlations between the
particles at $\textbf{r}_{1}$ and $\textbf{r}_{2}$, and in most contexts
this would represent a crude and generally inaccurate approximation.
However for \emph{slowly varying} $u_{1}$ we see that this particular
use of the (mean field) approximation can be very accurate, even at high
density. Choosing the constant of integration so that the bulk densities
in zero field satisfy $\rho^{B}=\rho_{0}^{B}$ we arrive at the simple
\emph{local molecular field} (LMF) equation
\cite{ katsov.k.weeks.jd:incorporating,weeks.jd:connecting} for
the effective field $\phi_{R}$:
\begin{equation}
\phi_{R}({\textbf{r}}_{1})=\phi({\textbf{r}}_{1})+\int d{\textbf{r}}_{2}
[\rho_{0}({\textbf{r}}_{2};[\phi_{R}])-\rho_{0}^{B}]u_{1}({\textbf{r}}_{12}),
\label{eq:lmf}
\end{equation}
which is the starting point for our work on nonuniform fluids with
both short and long ranged interactions. 

To solve this self consistent equation we need to determine the nonuniform
density $\rho_{0}({\textbf{r}}_{2};[\phi_{R}])$ in the presence of
the effective field $\phi_{R}$. The LMF approach does not specify
or require a particular way to do this. However since the intermolecular
interactions and the effective field have shorter ranges in the reference
system, both theory and simulations of the nonuniform structure are
usually easier to carry out than in the full system. Equation \ref{eq:rho0rho}
then provides the fundamental link between structure in the nonuniform
reference and full systems, and from this thermodynamic functions
can be determined. 

The name ``local molecular field'' is suggested by the direct analogy to the
spatially varying effective field introduced in the usual mean or molecular
field theory for a nonuniform Ising model.\cite{binderh,weekspair}
However, while the latter
theory is usually viewed as a crude approximation, the derivation
sketched above suggests that if a proper choice of a slowly-varying
$u_{1}$ is made, then accurate results should be found from a self-consistent
solution of the LMF equation in many cases, provided that an accurate treatment
of the density response
$\rho_{0}({\textbf{r}};[\phi_{R}])$ to a given $\phi_{R}$ is used.
In this paper we use the exact eq \ref{eq:lowrho} below
at low densities and results of computer simulations at higher densities,
so whatever errors remain arise only from the LMF approximation itself.
 
\subsection{Uniform systems}
\label{sec:uniform}

LMF theory is equally useful for uniform fluids.
\cite{vollmayr-lee.k.katsov.k.ea:using,weeks.jd.vollmayr.k.ea:intermolecular}
In particular eq
\ref{eq:lmf} is consistent with the physical idea that in a \emph{dense}
uniform fluid with $\phi=0$ the forces associated with the slowly
varying $u_{1}$ from oppositely situated particles essentially \emph{cancel}
in most relevant configurations.\cite{widomsci,weeks.jd.chandler.d.ea:perturbation}
Moreover, any residual effects are strongly
damped by the small compressibility at high density, so we expect that
the radial distribution functions will satisfy
\begin{equation}
g_{0}(r)\simeq g(r),\label{eq:gog}
\end{equation}
as predicted by eqs \ref{eq:crho0crho} and \ref{eq:lmf} for $\phi=\phi_{R}=0$.
For example, eq \ref{eq:gog} holds to a rather good approximation at high density
in the uniform Lennard-Jones (LJ) fluid provided that the WCA separation
\cite{chandler.d.weeks.jd:equilibrium,weeks.jd.chandler.d.ea:perturbation,weeks.jd.chandler.d.ea:role}
with its relatively slowly varying $u_{1}$ is used, showing the consistency
of the physical picture. 

This perfect cancellation argument can at best give reasonable results
only for uniform fluids at high density. However, LMF theory can be
applied to a general nonuniform fluid,
and by taking such a perspective and using only the basic eq \ref{eq:rho0rho},
we can significantly improve on the predictions of eq \ref{eq:gog}
for pair correlations in uniform fluids.
\cite{vollmayr-lee.k.katsov.k.ea:using,weeks.jd.vollmayr.k.ea:intermolecular}

Corrections to eq \ref{eq:gog} can be found by considering the
particular external field arising from a fluid particle fixed at the
origin, $\phi({\textbf{r}})=w(r)$. The induced density now gives
\begin{equation}
\rho(r;[w])=\rho^{B}g(r).\label{eq:rhog}\end{equation}
This exact equation\cite{percusfix} relating the nonuniform singlet density induced
by a fixed particle to the radial distribution function $g(r)$ in
the uniform fluid plays a key role in the theory below. There are
now net unbalanced forces arising from the fixed particle and eq
\ref{eq:lmf} predicts a nonzero $\phi_{R}$, which can be used
in eq \ref{eq:rho0rho} to give a more accurate approximation for
$\rho(r;[w])$ and hence $g(r)$.

For the uniform LJ fluid this approach accurately determines the small
corrections to eq \ref{eq:gog} at high
density.\cite{vollmayr-lee.k.katsov.k.ea:using,weeks.jd.vollmayr.k.ea:intermolecular}
Moreover at very low densities where eq \ref{eq:gog} would be
very inaccurate, eq \ref{eq:lmf} gives $\phi_{R}({\textbf{r}})=\phi(r)=w(r)$
and we obtain the \emph{exact} low density limit\cite{hansenmac} for
$g(r)=\exp[-\beta w(r)]$ by using eq \ref{eq:rho0rho} and the
exact low density limit of the reference system in the field $\phi_{R}$:
\begin{equation}
\rho_{0}({\textbf{r}};[\phi_{R}])=\rho^{B}e^{-\beta\phi_{R}({\textbf{r}})}.\label{eq:lowrho}
\end{equation}
 Here $\beta=(k_{B}T)^{-1}$. 

Similar accurate results for the nonuniform LJ fluid have been found
for more general external fields representing hard core solutes of
various sizes, for the liquid-vapor interface, and for drying
transitions.\cite{katsov.k.weeks.jd:incorporating,katsov.k.weeks.jd:on}
Thus LMF theory has provided a qualitatively and often quantitatively
accurate description of structure, thermodynamics, and phase transitions
in fluids with short ranged interactions.\cite{weeks.jd:connecting} 

\section{One Component Charged Hard Spheres}

\label{sec:OCCHS}In this paper we focus on the uniform OCCHS, where
there are $N$ positive ions in a volume $V$ with a uniform neutralizing
background that also penetrates the ions. The pair potential $w(r)$
for the ions in the OCCHS is usually written as
\begin{equation}
w(r)=w_{d}(r)+w_{q}(r),\label{eq:wd+wq}
\end{equation}
the sum of a hard sphere potential
\begin{equation}
w_{d}(r)\equiv\left\{ \begin{array}{c}
\infty,\,\, r\leq d\\
0,\,\, r>d\end{array}\right.\label{eq:wd}
\end{equation}
and the pair potential $w_{q}(r)$ arising from point charges of
magnitude $q$, where
\begin{equation}
w_{q}(r)\equiv{\displaystyle {\frac{q^{2}}{\epsilon r}}}.\label{eq:wq}
\end{equation}
The separation in eq \ref{eq:wd+wq} is a special case of eq
\ref{eq:wdiv} and more general separations of $w$ will prove useful
in the LMF theory developed below. In eq \ref{eq:wq} the solvent
is crudely represented by a uniform static dielectric constant $\epsilon$.
In the limit $d=0$, the OCCHS reduces to the OCP. 

It is convenient to introduce a characteristic length describing the
typical distance between neighboring particles. A standard choice
is the \emph{ion sphere radius} $a$ chosen so that
\begin{equation}
\frac{4\pi}{3}\rho^{B}a^{3}=1\label{eq:adef}
\end{equation}
The nearest neighbor spacing is about $1.6a$ when the ions are arranged
in a simple cubic lattice. 

Thermodynamic properties in the OCCHS can then be characterized using
two dimensionless parameters, the \emph{ionic strength}\begin{equation}
\Gamma\equiv\frac{\beta q^{2}}{\epsilon a},\label{eq:gammadef}\end{equation}
 which compares the bare Coulomb interaction energy between two ions
separated by the characteristic distance $a$ to $k_{B}T$, and the
\emph{hard sphere packing fraction}\begin{equation}
\eta\equiv\pi\rho^{B}d^{3}/6.\label{eq:etadef}\end{equation}
 Note that $d/a=2\eta^{1/3}.$ In the OCP $d=\eta=0$ and thermodynamic
properties depend only on the single dimensionless parameter $\Gamma.$

Pair correlations between the ions in the uniform fluid are most conveniently
described in terms of the density change induced by fixing a particle
at the origin: \begin{equation}
\Delta\rho(r;[w])\equiv\rho(r;[w])-\rho^{B}=\rho^{B}h(r),\label{eq:hdef}\end{equation}
 where $\rho^{B}=N/V$ and $h(r)\equiv g(r)-1$ is the pair correlation
function in the uniform fluid. 

The unique consequences of the long ranged interaction in the OCCHS
are most easily seen by taking the Fourier transform of eq \ref{eq:hdef}
and defining the dimensionless \emph{structure factor}\begin{equation}
S(k)\equiv1+\Delta\hat{\rho}(k;[w])=1+\rho^{B}\hat{h}(k),\label{eq:S(k)def}\end{equation}
 where the caret denotes a Fourier transform. As argued generally
in the seminal work of Stillinger and Lovett\cite{stillinger.fhj.lovett.r:general} (SL),
there should be complete screening at long wavelengths of any induced
charge distribution in a conducting ionic fluid. This constrains the
behavior at small wavevectors of the charge-charge correlation function.
For the OCCHS the only nontrivial correlations are between the positive
ions and the results of SL reduce to the requirement that $S(k)$
has the \emph{universal} form \begin{equation}
S(k)=0+k^{2}/k_{D}^{2}+{\textrm{O}}(k^{4}),\label{eq:SSL}\end{equation}
independent of any details of the short ranged core potential $w_{d}$
or any other short ranged interactions that might exist. Here $k_{D}$
is the \emph{Debye wavevector}, defined by
\begin{equation}
k_{D}^{2}\equiv4\pi\beta q^{2}\rho^{B}/\epsilon=3\Gamma/a^{2}.\label{eq:kD}
\end{equation}

The exact vanishing of $S(k)$ at $k=0$ arises from electrical neutrality
(the {}``zeroth'' moment condition) and the fixed coefficient of
the quadratic term is an example of the famous SL \emph{second moment
condition}.\cite{stillinger.fhj.lovett.r:general,outhwaite.cw:comment}
This behavior is distinctly different than that found in any fluid
with short ranged interactions, where $S_{0}(k)$ at $k=0$ is finite,
proportional to the compressibility, and the coefficient of $k^{2}$
depends on the details of the intermolecular interactions. 

Thermodynamic properties can be found by integration of the correlation
functions. In particular, accounting for the background by taking
the appropriate limits of the standard result for a two component
system,\cite{palmerweeks} the excess internal energy (over the ideal gas) of the uniform
OCCHS can be exactly written as
\begin{equation}
\frac{\beta E^{ex}}{N}=\frac{\beta\rho^{B}}{2}\int d{\textbf{r}}
{\displaystyle {\frac{q^{2}}{\epsilon r}}}h(r).\label{eq:energy}
\end{equation}

\section{LMF theory for OCCHS}

\label{sec:lmfocchs}

\subsection{Gaussian charge distribution}

\label{sec:gaussian}

We want to use the general LMF equation to describe the ion correlation
functions in the uniform OCCHS. It is clear from the derivation in
section \ref{sec:lmf} that a proper separation of the interaction potential
$w=u_{0}+u_{1}$ is required for this self-consistent approach to
be accurate. At first glance it may seem natural to use the separation
on the right side of eq \ref{eq:wd+wq}, where $u_{0}$ is taken
to be the hard core potential $w_{d}$ and $u_{1}$ is the full point
charge interaction $w_{q}$ in eq \ref{eq:wq}. However at short
distances outside the hard core the Coulomb potential can be strong
and rapidly varying and such interactions cannot be accurately treated
by the averaging used in eq \ref{eq:lmf}. 

Indeed in the OCP with no hard core there are arbitrarily large and
rapidly varying interactions as $r\rightarrow0$. This limit makes
it clear that we should try to separate the point charge Coulomb pair
interaction $w_{q}(r)$ itself into a slowly varying part $w_{q1}(r)$,
which we will take as a particularly appropriate $u_{1}(r)$ to use
in the LMF theory, and then combine the remainder $w_{q0}(r)\equiv w_{q}(r)-u_{1}(r)$
with $w_{d}(r)$ (and more generally with any other short ranged core
interactions that exist) to give the associated $u_{0}(r)$. 

This strategy differs from that used in many density functional
and integral equation methods, where one first chooses a mathematically
convenient or especially simple \emph{reference potential} $\tilde{u}_{0}(r)$
and then treats the remainder $w(r)-\tilde{u}_{0}(r)$ as a perturbation,
taking advantage of the particular form of the reference system in
whatever approximate theories are used. However, for Coulomb systems
at least, the existing theories often have large and uncontrolled
errors with the usual choices of $\tilde{u}_{0}$. We believe our
approach of choosing a slowly varying $u_{1}$ for use in the LMF theory offers
many conceptual and computational advantages, and it connects directly
to similar physically motivated work on fluids with short ranged interactions. 

To that end we interpret the $1/r$ term in $w_{q}$ as the electrostatic
potential of a unit point charge in vacuum. The same slowly varying
asymptotic behavior would come from any other normalized charge distribution
and the ``smearing'' of the point charge would produce a less
rapidly varying potential at small $r$. This suggests using a properly chosen
charge distribution to determine $w_{q1}(r)$ or $u_1(r)$.

Consider in particular as
in the Ewald sum method \cite{hansenmac,frenkelsmit} a normalized unit
\emph{Gaussian charge distribution}
\begin{equation}
P_{\sigma}(r)=\pi^{-3/2}\sigma^{-3}\exp[-(r/\sigma)^{2}],\label{eq:Psigmar}
\end{equation}
where $\sigma$ indicates the length scale of the smearing. The particular
advantages of this choice will soon become apparent. Our use
of a Gaussian charge distribution to determine a slowly varying part of
the Coulomb pair interaction is simpler than in the Ewald sum method, which
considers periodic images of ion configurations with embedded screening
(negative) and compensating (positive) Gaussian charge distributions.
\cite{frenkelsmit} 

The electrostatic potential $v_{\sigma}(r)$ arising from eq \ref{eq:Psigmar}
satisfies Poisson's equation
$\nabla^{2}v_{\sigma}(r)=-4\pi P_{\sigma}(r)$, which is easily solved
by Fourier transform to give
\begin{equation}
\hat{v}_{\sigma}(k)=\frac{4\pi}{k^{2}}\exp[-{\textstyle {\frac{1}{4}}}(k\sigma)^{2}],
\label{eq:vsigmak}
\end{equation}
or in $r$-space,
\begin{equation}
v_{\sigma}(r)=\frac{1}{r}\mathop{\textrm{erf}}(r/\sigma),\label{eq:vsigmar}
\end{equation}
where erf is the usual error function.\cite{frenkelsmit} The point
charge model corresponds to the limit $\sigma=0$.

Thus we can write
\begin{equation}
\frac{1}{r}=\frac{1}{r}{\textrm{erfc}}(r/\sigma)
+\frac{1}{r}\mathop{\textrm{erf}}(r/\sigma),\label{eq:1/r}
\end{equation}
and use this identity to extract from the dimensionless Coulomb pair
interaction $\beta w_{q}(r)$ a $\sigma$-dependent perturbation piece
$\beta u_{1}(r)$:
\begin{equation}
\beta u_{1}(r)=\frac{\beta q^{2}}{\epsilon}v_{\sigma}(r)
=\frac{\beta q^{2}}{\epsilon r}\mathop{\textrm{erf}}(r/\sigma).
\label{eq:u1sigma}
\end{equation}
This perturbation remains finite as $r\rightarrow0,$ with
$\beta u_{1}(0)=2\pi^{-1/2}\beta q^{2}/(\epsilon\sigma)$.

As illustrated in Figure \ref{cap:potsept},
with appropriate choices of $\sigma$ we can produce a very slowly
varying $u_{1}(r),$ which from eq \ref{eq:vsigmak} also decays
very rapidly in $k$-space. As argued in section \ref{sec:lmf}, these
are the qualitative features that would be most appropriate for a
perturbation $u_{1}$ to give accurate results from LMF theory. The
choice of $\sigma$ in the Gaussian charge distribution permits a
\emph{controlled use} of the LMF approximation, and as shown below,
with proper choices of $\sigma$ the LMF theory can give exceptionally
accurate results. 

Any particular choice of $\sigma$ in eq \ref{eq:u1sigma} then
fixes the associated reference system interaction\cite{ceperley} as
\begin{equation}
\beta u_{0}(r)=\beta u_{d}(r)+\frac{\beta q^{2}}{\epsilon r}
{\textrm{erfc}}(r/\sigma).\label{eq:u0sigma}
\end{equation}
The Coulomb part $w_{q0}(r)$ of the reference interaction decays
very rapidly for $r>\sigma.$ For large $r$ we have
\begin{equation}
\beta w_{q0}(r)\equiv\frac{\beta q^{2}}{\epsilon r}
{\textrm{erfc}}(r/\sigma)\sim\frac{\beta\sigma q^{2}}{\sqrt{\pi}\epsilon r^{2}}
\exp[-(r/\sigma)^{2}].\label{eq:u0asymp}
\end{equation}

We call the special reference systems that result from optimal
choices of $\sigma$ as discussed in section \ref{sec:choiceofsigma} below
\emph{mimic systems}, since at high density
the local structure in the uniform mimic system as exhibited in $g_{0}(r)$
very accurately approximates the $g(r)$ of the full system as in
eq \ref{eq:gog}. This again illustrates the consistency and accuracy
of the LMF approach when an appropriate mimic system is used. 

Equation \ref{eq:u1sigma} can also be interpreted physically as
the Coulomb energy arising from two ions each with a rigid Gaussian
charge distribution, eq \ref{eq:Psigmar},
with a width $\tilde{\sigma}=\sigma/\sqrt{2}$.
More generally, in ionic solutions we can always replace point charges
on the ions by rigid charge distributions without changing any physics
if we appropriately modify the core interactions as in eq \ref{eq:u0sigma}.
This can be very useful because the rapidly varying short ranged parts of the
Coulomb interaction can often be more accurately treated by the same
specialized methods used for the other strong core interactions, which must be
present in any realistic model of ionic solutions.

\subsection{Scaled LMF equation for the OCCHS}

\label{sec:scaledLMF}

We now apply the general LMF eq \ref{eq:lmf} to the OCCHS
in the special case where the external field $\phi({\textbf{r}})=w(r)$
is that resulting from an ion fixed at the origin, given by eq \ref{eq:wd+wq}.
This choice allows us to describe uniform fluids, as discussed in Section
\ref{sec:uniform}. We take advantage of spherical
symmetry and use the Gaussian charge
separation of $w(r)$ given in eqs \ref{eq:u1sigma} and \ref{eq:u0sigma}. 

The LMF eq \ref{eq:lmf} can be naturally rewritten in terms of the
more slowly varying ``perturbation part'' of the effective field
\begin{equation}
\phi_{R1}(r)\equiv\phi_{R}(r)-u_{0}(r).\label{eq:phir1r}
\end{equation}
If the perfect cancellation argument were exact, then $\phi_{R}(r)=u_{0}(r)$,
or $\phi_{R1}(r)=0,$ corresponding to a fixed mimic particle at the
origin, and the resulting induced density in the mimic system would
be $\Delta\rho_{0}(r;[u_{0}])=\rho^{B}h_{0}(r)$, with $h_{0}(r)$
the pair correlation function in the uniform mimic system. The LMF
equation corrects this approximation by determining a finite short
ranged effective field perturbation $\phi_{R1}(r),$ which we can
picture as arising from a modified solute particle at the origin,
\cite{vollmayr-lee.k.katsov.k.ea:using,weeks.jd.vollmayr.k.ea:intermolecular}
that takes better account of the locally averaged effects of the slowly
varying interactions $u_1$. 

Taking Fourier transforms, and introducing for reasons that will soon
become apparent a multiplicative parameter $\alpha$ that scales the
amplitude of $\hat{\phi}_{R1}$, a scaled version of the LMF equation
can be written as
\begin{equation}
\beta\rho^{B}\hat{\phi}_{R1}(k)=\frac{\alpha k_{D}^{2}}{k^{2}}
\exp[-{\textstyle {\frac{1}{4}}}(k\sigma)^{2}]S_{R}(k),
\label{eq:lmfalpha}
\end{equation}
where \begin{equation}
S_{R}(k)\equiv1+\Delta\hat{\rho}_{0}(k;[\phi_{R}])\label{eq:SR(k)}\end{equation}
For now we simply note that the original unscaled LMF equation has
$\alpha=1,$ and view $\alpha$ as a parameter at our disposal.
A similar scaling of the LMF equation for systems with short ranged interactions
was discussed earlier.\cite{katsov.k.weeks.jd:incorporating}

Before giving quantitative results in sections \ref{sec:lowrhoresults}
and \ref{sec:highdensity} below, let us discuss some qualitative
features of a self-consistent solution of eq \ref{eq:lmfalpha}.
Such a solution would determine a short ranged effective field, implying
a $\hat{\phi}_{R1}(k)$ that is finite as $k\rightarrow0,$ along
with the associated $\Delta\hat{\rho}_{0}(k;[\phi_{R}]).$ By the
fundamental assumption of LMF theory in eq \ref{eq:rho0rho}, the
latter is the LMF approximation to the full $\Delta\hat{\rho}(k;[w])$
in the OCCHS. In particular $S_{R}(k)$ in eq \ref{eq:SR(k)} is
the LMF approximation to $S(k)$ in eq \ref{eq:S(k)def}, and when
no confusion will result, we will simply write $S(k)$. $S_{R}(k)$
should be carefully distinguished from $S_{0}(k)\equiv1+\Delta\hat{\rho}_{0}(k;[u_{0}])$,
which equals the structure factor in the \emph{uniform} mimic system
with $\phi_{R1}=0.$

\subsection{Choice of $\alpha$}

In order that $\hat{\phi}_{R1}(k)$ remain finite as $k\rightarrow0$
in eq \ref{eq:lmfalpha}, it is clear that with any choice of $\alpha,$
the associated $S_{R}(0)$ must vanish identically. (In practice it
is not easy to ensure this in a self consistent iterative solution
of eq \ref{eq:lmfalpha}, and we give in the Appendix in eq \ref{eq:lmfK}
an alternate but equivalent version that is numerically more stable.)
Thus the LMF theory gives an approximate structure factor $S_{R}(k)$
that always obeys the exact neutrality condition. Its expansion at
small $k$ has the general form
\begin{equation}
S_{R}(k)=0+B(\alpha)k^{2}+{\textrm{O}}(k^{4}),\label{eq:SR(k)expand}
\end{equation}
resembling eq \ref{eq:SSL}, but the coefficient $B(\alpha)$
of $k^{2}$ depends on $\alpha$ and does not necessarily obey the
exact second moment condition. However, by substituting eq \ref{eq:SR(k)expand}
into eq \ref{eq:lmfalpha}, we see that the exact result $B(\alpha)=k_{D}^{-2}$
is found if $\alpha$ is chosen self consistently so that\cite{potentialnorm}
\begin{equation}
\beta\rho^{B}\hat{\phi}_{R1}(0)=\alpha.\label{eq:alpha2nd}\end{equation}

Thus by proper choice of $\alpha$, we can guarantee that the approximate
structure factor $S_{R}(k)$ also obeys the exact second moment condition.
We show below in section \ref{sec:lowrhoresults} that with optimal
choices of the key parameter $\sigma$, even the unscaled LMF theory
with $\alpha=1$ often gives very good numerical results. However
it is conceptually important to realize that the LMF approach can
be naturally generalized as in eq \ref{eq:lmfalpha} so that the
exact second moment condition is satisfied, and this adds essentially no
numerical costs to the self consistent solution.
We use eq \ref{eq:lmfalpha} along with
eq \ref{eq:alpha2nd} in most of the work reported below, and usually
refer to this generalized approach as the LMF theory. If we want to
emphasize that the second moment condition is satisfied, we will refer
to the LMF2 theory, and distinguish this from the original unscaled
version, which we will call the LMF0 theory. 

\subsection{Choice of $\sigma$}
\label{sec:choiceofsigma}
The ability to choose a $\sigma$ larger than some $\sigma_{\min}$
allows for a consistent application of LMF theory to ionic fluids,
ensuring that the LMF approximation is used only for slowly varying
parts of the Coulomb interactions. The choice of $\sigma_{\min}$ determines
an effective Coulomb core size from the core component $w_{q0}(r)$
of the Coulomb interaction. This may be larger or smaller than the
``physical'' core size $d$, which can be varied independently
in the OCCHS. 

For strong coupling with $\Gamma\gg1$, we expect
considerable cancellation of the very strong forces from ions at distances
larger than the nearest neighbor spacing $a$. Thus the effective
core size $\sigma_{\min}/a$ should be of order unity and essentially
independent of $\Gamma$ at large $\Gamma$.  

However, for weak couplings with $\Gamma\ll1$ we would expect
that any choice of $\sigma\gtrsim \sigma_{\min}$ with
$\sigma_{\min}\simeq \Gamma a$ will be sufficient for the LMF equation
to give good results. This (conservative) choice of $\sigma_{\min}$
corresponds to the Bjerrum length.\cite{hansenmac} Only on length scales
$\sigma_{\min}$ much less that the average neighbor separation $a$
will even the bare Coulomb interactions between ions exceed $k_{B}T$,
which would characterize an effective Coulomb core size. A more detailed
argument \cite{sigmachoice} suggests that we can take even smaller $\sigma$,
including $\sigma =0$, for $\Gamma\ll1$ and still get accurate results from the
LMF theory. We will see below that these qualitative considerations hold
true generally.

In particular, by choosing $\sigma$ large enough we can guarantee
that $\beta\rho^{B}\hat{\phi}_{R1}(k)$ is nonzero only at small wavevectors,
since the Gaussian factor in eq \ref{eq:lmfalpha} causes rapid
decay for $k\sigma\gtrsim2$. The Gaussian charge distribution produces
this very efficient localization of $\beta\rho^{B}\hat{\phi}_{R1}(k)$
to small wavevectors, and is much superior in this regard to most
other smooth distributions. 

This property is very important at high density and strong
coupling where $S_{R}(k)$ can have significant structure at $ka\simeq2\pi,$
where $a$ roughly measures the typical distance between nearest neighbor
particles. At those larger wavevectors
that characterize short ranged structure in $r$-space,
$\beta\rho^{B}\hat{\phi}_{R1}(k)$
essentially vanishes for any choice of $\sigma\geq$ $\sigma_{\min}\simeq$
$a.$ Thus for such wavevectors we have $S_{R}(k)\simeq S_{0}(k)$ from a
crude linear response argument. Differences in these functions should show
up only at small $k$, where $S_{R}(k)$ will satisfy the SL moment
conditions while $S_{0}(k)$ remains finite as $k\rightarrow0.$ However
at high densities the compressibility in the mimic system is small,
so that the absolute differences between $S_{R}(k)$ and $S_{0}(k)$
remain small even at small wavevectors, as will be illustrated in
Figure \ref{cap:occhsg20.sk} below. 

On taking inverse transforms we then expect that $h_{0}(r)\simeq h(r)$
holds true to a very good approximation at high densities, as was
qualitatively suggested by the cancellation argument leading to eq
\ref{eq:gog}. Thus we predict a \emph{family} of uniform mimic
systems for different $\sigma\geq$ $\sigma_{\min},$ all of which
should give essentially the same short ranged structure at high density
that closely approximates that of the full ionic system. This is a
dramatic example showing that the inverse problem of uniquely determining
the intermolecular potential from $h_{0}(r)$ can be ill-conditioned.

A ``molecular-sized'' choice of $\sigma\simeq \sigma_{\min}$
for the mimic system is considerably smaller than the typical choices
made in Ewald sum methods, where $\sigma$ is usually taken to be proportional
to the simulation system size.\cite{frenkelsmit} Larger $\sigma$ values
will give equally good results, provided that
the resulting mimic system is described accurately. However generally
there is little to gain from such choices, since the LMF theory is already
consistent for $\sigma\simeq \sigma_{\min}$, and it
may be more difficult to treat the longer ranged
interactions in mimic systems with larger $\sigma$.
Thus optimal choices for $\sigma$ are generally
near $\sigma_{\min}$.

At lower densities and weak coupling there is little structure in
$S_{R}(k)$ and $S_{0}(k)$ at larger wavevectors, and
we can choose a much smaller $\sigma_{\min}$ as
argued above and still make consistent use of the LMF approximation.
Very accurate results for $S_{R}(k)$ are again found from a self-consistent
solution of eq \ref{eq:lmfalpha}, but the long wavelength perturbations
from $\beta\rho^{B}\hat{\phi}_{R1}(k)$ are not damped by low compressibility
in the mimic system, and $h_{0}(r)$ can differ noticeably from $h(r).$
As $\Gamma\rightarrow0,$ our theory reduces correctly to the exact
Debye-H\"{u}ckel (DH) limit.\cite{hansenmac}

\section{Results at Low Density: Mimic Poisson-Boltzmann Approximation}

\label{sec:lowrhoresults}

At low enough bulk densities, the mimic system's response to $\phi_{R}(r)$
can be described using the simple ideal gas Boltzmann factor as in
eq \ref{eq:lowrho}, so that
\begin{equation}
\Delta\rho_{0}({\textbf{r}};[\phi_{R}])=\rho^{B}[e^{-\beta\phi_{R}({\textbf{r}})}-1].
\label{eq:hbz}
\end{equation}
This also represents the LMF prediction for the full system's $\rho^{B}h(r)$,
and requires only that second and higher order virial corrections
to the \emph{mimic system's} pair correlation function can be ignored.
When eq \ref{eq:hbz} is substituted in the LMF equation \ref{eq:lmfalpha},
a closed equation for $\phi_{R}({\textbf{r}})$ results. A self consistent
solution is readily found by iteration, using the equivalent but numerically
more stable version of the LMF equation in eq \ref{eq:lmfK}. 

\begin{figure}
\includegraphics[%
  scale=0.32]{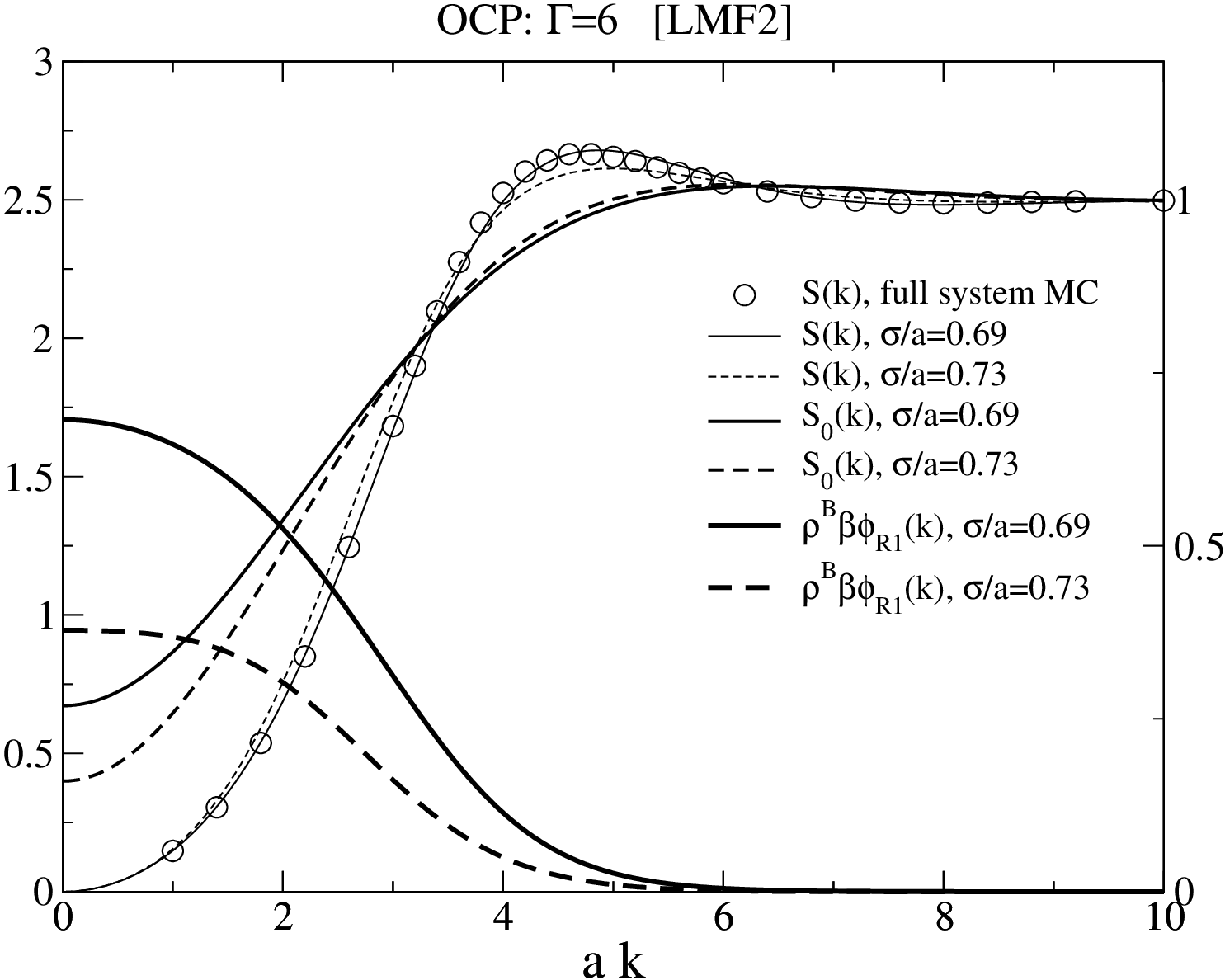}
\caption{\label{cap:ocpdiffsigma} $S_{R}(k)=S(k)$ at moderate coupling for the OCP
computed using the MPB theory with different $\sigma $'s and compared to
$S_{0}(k)$ for the uniform mimic system. Also shown are the associated
$\rho \beta \hat{\phi}_{R1}$, which use the scale on the left $y$-axis.
When $\rho \beta \hat{\phi}_{R1}$ is
taken into account using the MPB theory, both choices of $\sigma $ give very
similar $S(k)$ that compare well with simulation data for the full
system.\cite{galam}}
\end{figure}

Equation \ref{eq:hbz} is the same structural approximation that
is used in the Poisson-Boltzmann (PB) theory.\cite{hansenmac}
Indeed if we set $\alpha=1$
and $\sigma=0$ in the LMF equation \ref{eq:lmfalpha} combined with
eq \ref{eq:hbz}, the results reduce \emph{exactly} to those of
the standard PB theory. The PB theory thus results from taking the
full Coulomb interaction of eq \ref{eq:wq} as the perturbation
$u_{1}$ in the LMF equation and using the Boltzmann approximation
for the density response. 

We refer to the low density limit of our theory, where eq \ref{eq:hbz}
is used in eq \ref{eq:lmfalpha}, as the \emph{mimic Poisson-Boltzmann}
(MPB) theory. The MPB theory differs from the PB theory only by the choice
of $\sigma$ yielding a consistent mimic system along with
a choice of $\alpha$ that ensures that the second moment condition
is exactly satisfied. As we will see, these simple modifications greatly
improve the accuracy and range of validity of the MPB theory. 

\subsection{MPB theory for OCP}

Consider first the OCP, where there is no length scale in the potential
to suggest an intrinsic core size. A qualitative discussion of the
choice of the effective size $\sigma_{\min}$ was given above. In
practice it is easy to determine $\sigma_{\min}$ by solving eq \ref{eq:lmfalpha}
using successively larger values of $\sigma.$ For $\sigma<\sigma_{\min}$
the results vary strongly with $\sigma$ and are generally very inaccurate.
But for all $\sigma>\sigma_{\min}$ the LMF theory is consistent and
should give very similar predictions for the full system's structure
as exhibited in $S_{R}(k)$, even though the effective fields $\beta\rho^{B}\hat{\phi}_{R1}(k)$
and the uniform mimic systems' structure factors $S_{0}(k)$ can still
vary strongly with $\sigma.$ This is illustrated in Figure \ref{cap:ocpdiffsigma}
for the state with $\Gamma=6$, where the convergence of the results
for $\sigma/a=0.69$ and $\sigma/a=0.73$ is shown.

\begin{figure}
\includegraphics[%
  scale=0.32]{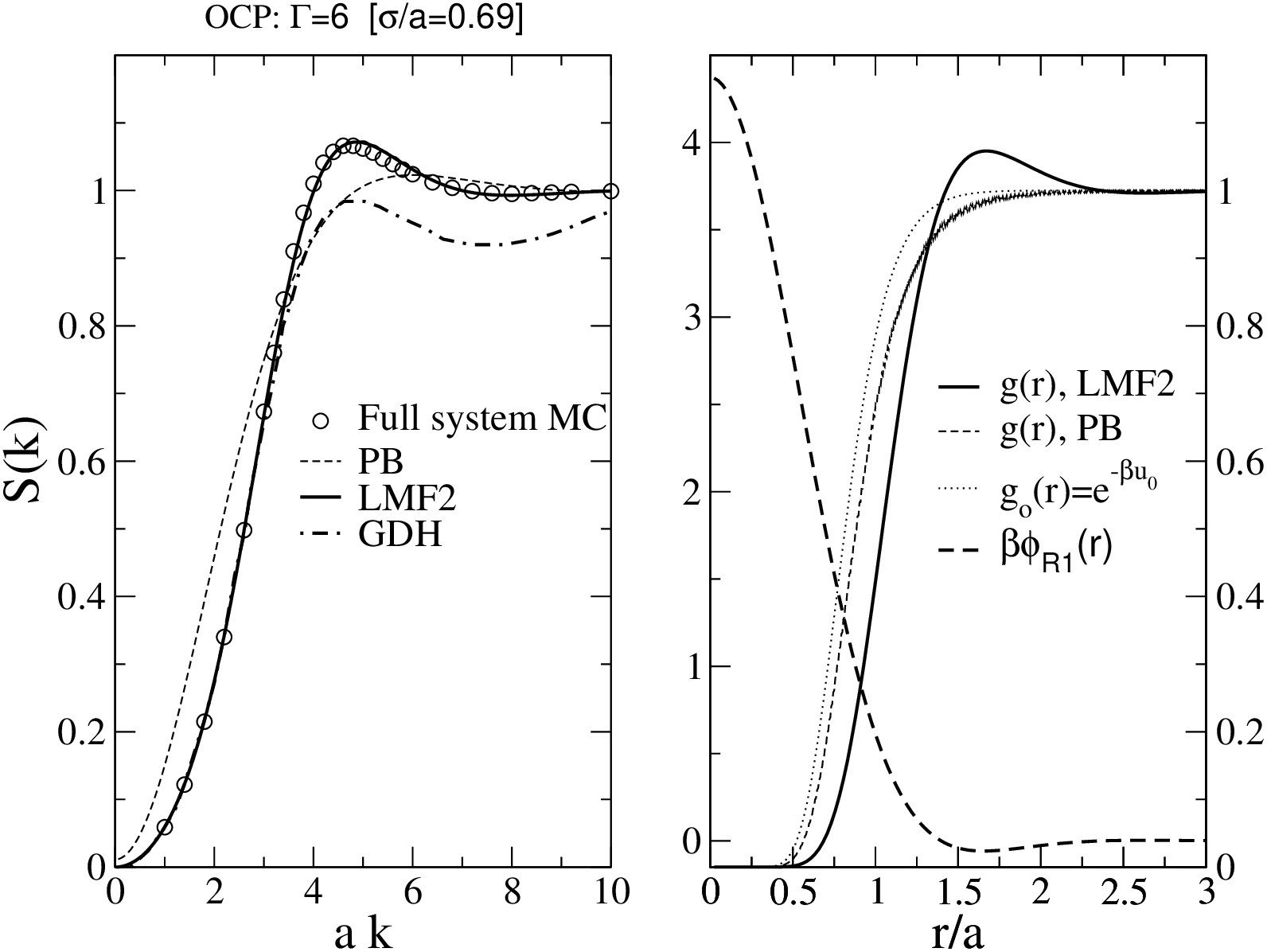}
\caption{\label{cap:ocp} Moderate coupling OCP structure computed using the MPB
theory. In the left graph, the LMF2 $S(k)$ is compared with the result of the
generalized Debye H\"{u}ckel theory, \cite{tamashiro.mn.levin.y.ea:one-component} (GDH)
and the usual PB theory. The right graph makes the same comparison for
$g(r)$ and also shows the effective field perturbation $\beta \phi _{R1}(r),$
which uses the scale on the left $y$-axis.  Both the LMF2 and GDH solutions
satisfy the second moment condition while PB does not. The GDH result is
expressed as an expansion and computed up to its $l=6$ term.\cite
{tamashiro.mn.levin.y.ea:one-component}}
\end{figure}

\begin{figure}
\includegraphics[%
  scale=0.32]{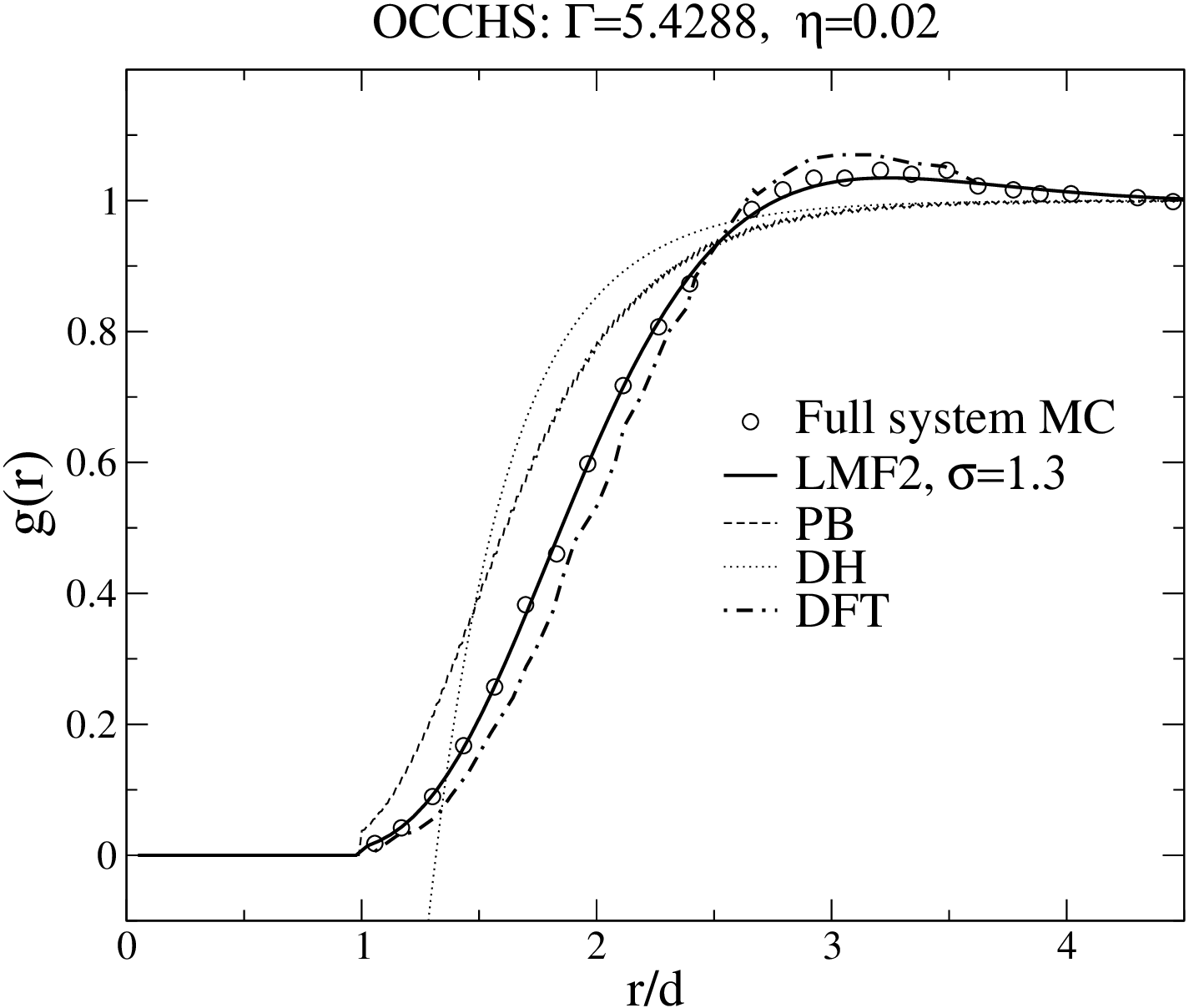}
\caption{\label{cap:occhs} Low density OCCHS structure. The LMF2 $g(r)$ from the MPB
theory is compared with the result of density functional
theory\cite{penfold.r.nordholm.s:simple} (DFT)
and the PB and DH approximations. Though
constrained to be zero inside the hard core by a boundary condition, the
DH $g(r)$ has a negative region near $r=d$, where $u_{0}(r)$ is also rapidly
rising, with a value of $-1.62$ at contact. The PB and DH theories fail to
capture qualitatively the onset of oscillation in $g(r)$ at this moderate
coupling strength.}
\end{figure}

\begin{figure}
\includegraphics[%
  scale=0.32]{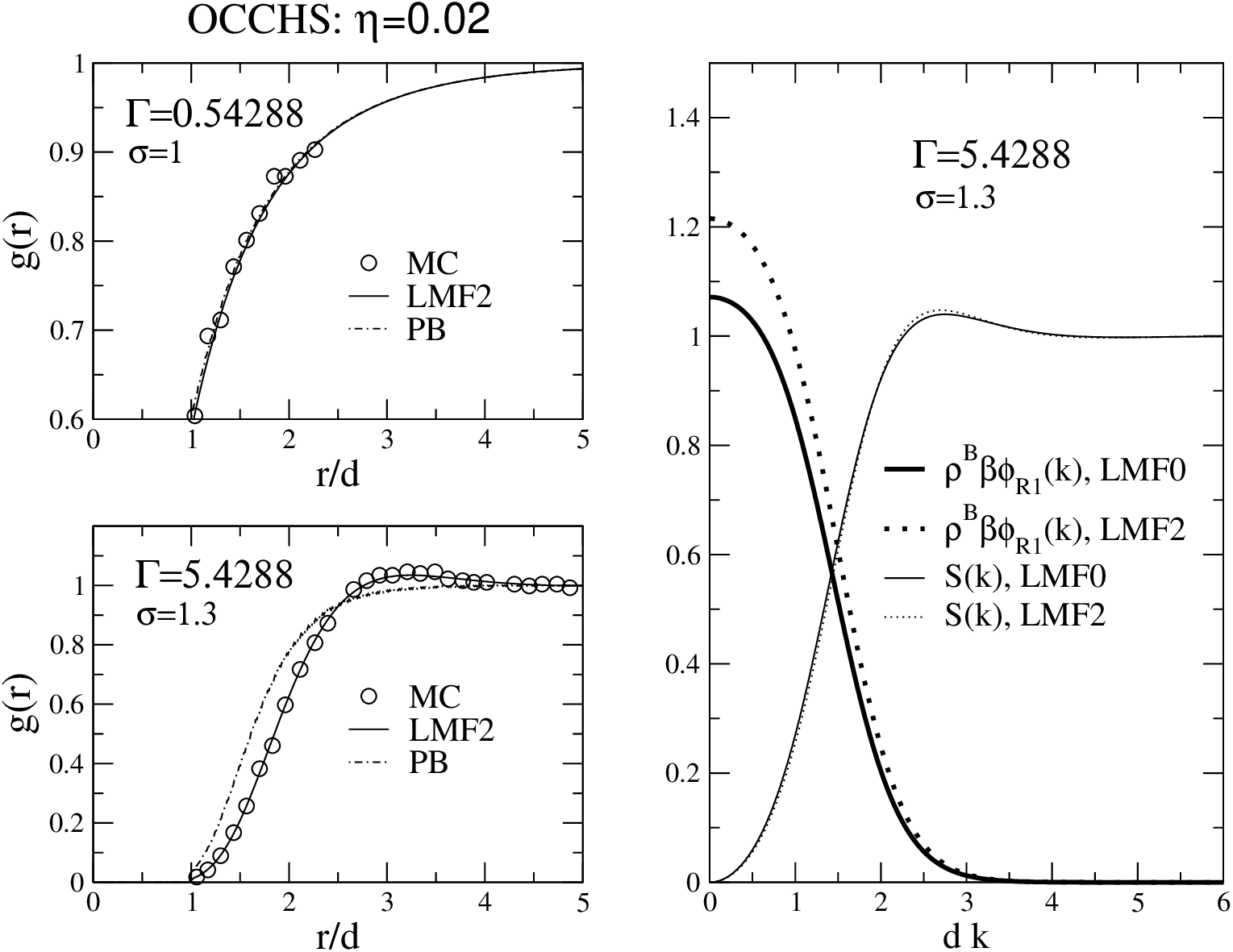}

\caption{\label{cap:occhspblim_2ndmntornot} The upper and lower left
graphs show the MPB
and PB approximations for the OCCHS $g(r)$ compared to
MC data\cite{penfold.r.nordholm.s:simple} for the full system for
weak and moderate ionic strengths. The PB approximation is satisfactory only
at weak couplings. The right graph shows that varying $\alpha $ so that
the second moment condition is satisfied in the MPB theory changes the
amplitude of $\beta \rho \hat{\phi}_{R1}(0)$ but in this case the effects
on $S(k)$ are hardly visible on the scale of the graph.}
\end{figure}

The LMF equation itself would continue to give (even more) accurate
results for larger $\sigma$ if an accurate theory for the structure
$\Delta\rho_{0}({\textbf{r}};[\phi_{R}])$ induced by a given $\phi_{R}$
is used. However the simple Boltzmann approximation used in eq \ref{eq:hbz}
for this quantity must fail at higher densities where there are significant
correlations between mimic system particles. This sets a $\sigma_{\max}$
above which the results of the MPB theory become inaccurate, very
roughly estimated by $\rho^{B}\sigma_{\max}^{3}\lesssim0.1$ as for
hard sphere fluids. 

As $\Gamma$ increases, eventually the $\sigma_{\min}$ needed for
the accuracy of the LMF approximation exceeds this $\sigma_{\max}$
and the MPB theory fails. The internal consistency or inconsistency
as $\sigma$ is varied is very evident from the MPB theory itself.
In practice we find very good and consistent results for all $\Gamma\lesssim6$
and the slight differences in the $S(k)$ curves in Figure \ref{cap:ocpdiffsigma}
and deteriorating results at larger $\sigma$ indicate that we are
near the upper limit of $\Gamma$ where the MPB theory can be trusted.
This represents a surprisingly strong coupling, since
the lowest order Boltzmann approximation for the structure in eq
\ref{eq:hbz} is used, and shows the virtues of choosing a
mimic system. 

As shown in Figure \ref{cap:ocp},
the results of the MPB theory for $\Gamma=6$ are in very good agreement
with computer simulations,\cite{galam} and compare very favorably to those of
the usual (nonlinear) PB theory or the generalized Debye-H\"{u}ckel
(GDH) theory developed by Levin and coworkers.\cite{tamashiro.mn.levin.y.ea:one-component}
Note that the MPB theory,
unlike the usual PB approximation, can predict oscillations in both
$S(k)$ and $g(r)$ from the self consistent determination of $\phi_{R1}(r)$
despite using only the lowest structural approximation, eq \ref{eq:hbz}. 

\subsection{MPB theory for OCCHS}

We now turn to the OCCHS. This has more complex structure because
of possible competition between correlations induced by the hard cores
and the soft repulsive Coulomb interactions. We follow the usual convention
where lengths are measured in terms of the hard core diameter $d$. 

In Figure \ref{cap:occhs}
we compare the results of the MPB theory to MC simulations, to results
of a density functional treatment, and to the PB and DH theories for
a low density state with $\eta=0.02$ but with a moderate ionic strength
$\Gamma=5.43.$ The MPB theory gives excellent results with a molecular-sized
$\sigma=1.3d$, noticeably better than those of the considerably more
complicated density functional theory.\cite{penfold.r.nordholm.s:simple} 

The right panel in Figure \ref{cap:occhspblim_2ndmntornot}
illustrates the role of $\alpha$ for this state, and shows that
the self consistent choice of $\alpha=1.2$ can ensure that the exact second
moment condition is satisfied, though on the scale of the graph the
differences between the $S_{R}(k)$ with $\alpha=1$ are hardly visible.
The left panels show that for much weaker coupling with $\Gamma=0.54$
even the usual PB approximation (with $\sigma=0$) gives good results, indicating that
at weak coupling the choice of $\sigma$ is not so important.
However, unlike the MPB theory, the PB approximation can satisfy the
second moment condition exactly only in the limit $\Gamma\rightarrow0,$
where it reduces to the DH approximation.\cite{hansenmac}

\section{Results at High Density and Strong Coupling}

\label{sec:highdensity}

\subsection{Structure in Mimic and Full Systems}

\begin{figure}
\includegraphics[%
  scale=0.32]{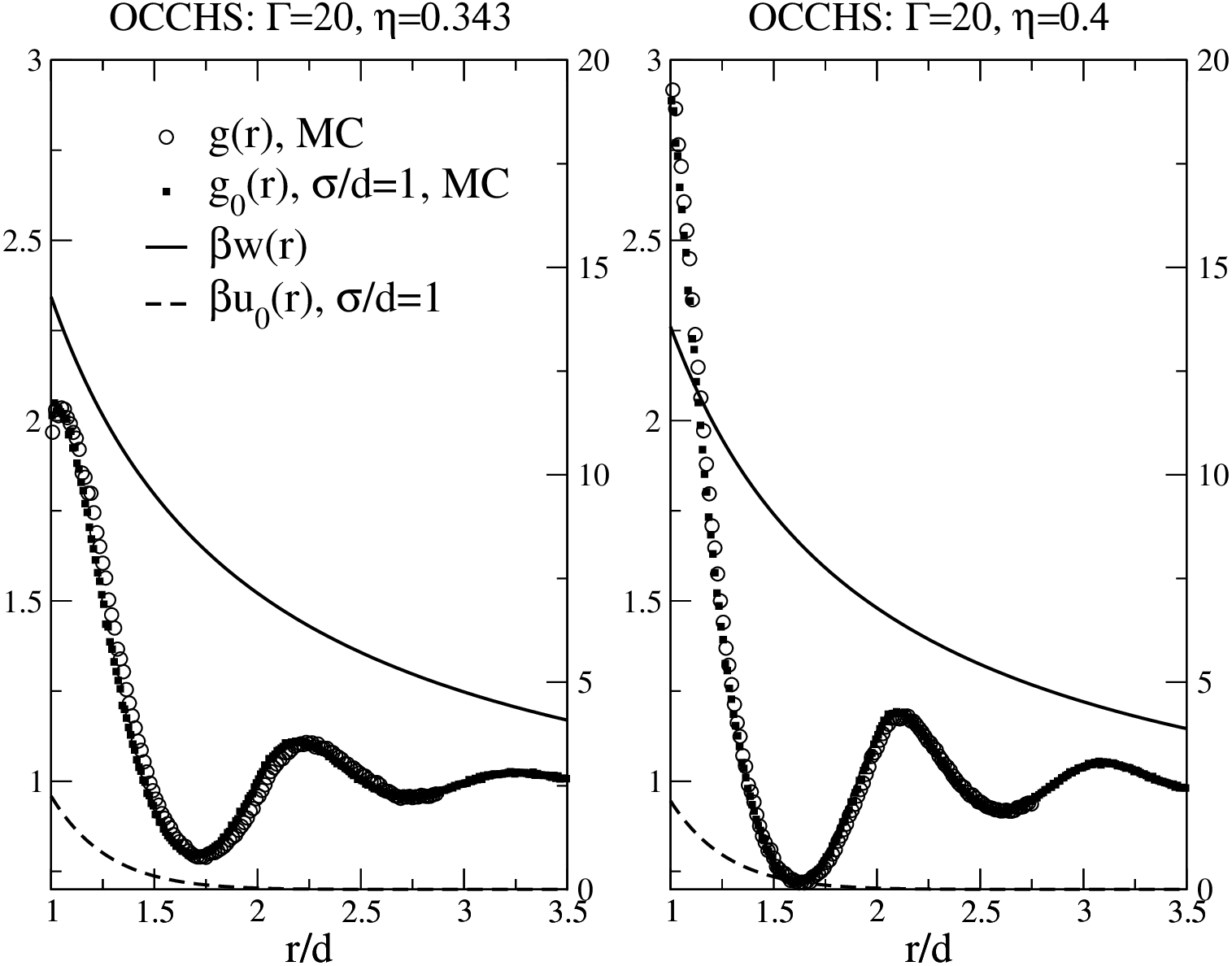}

\includegraphics[%
  scale=0.32]{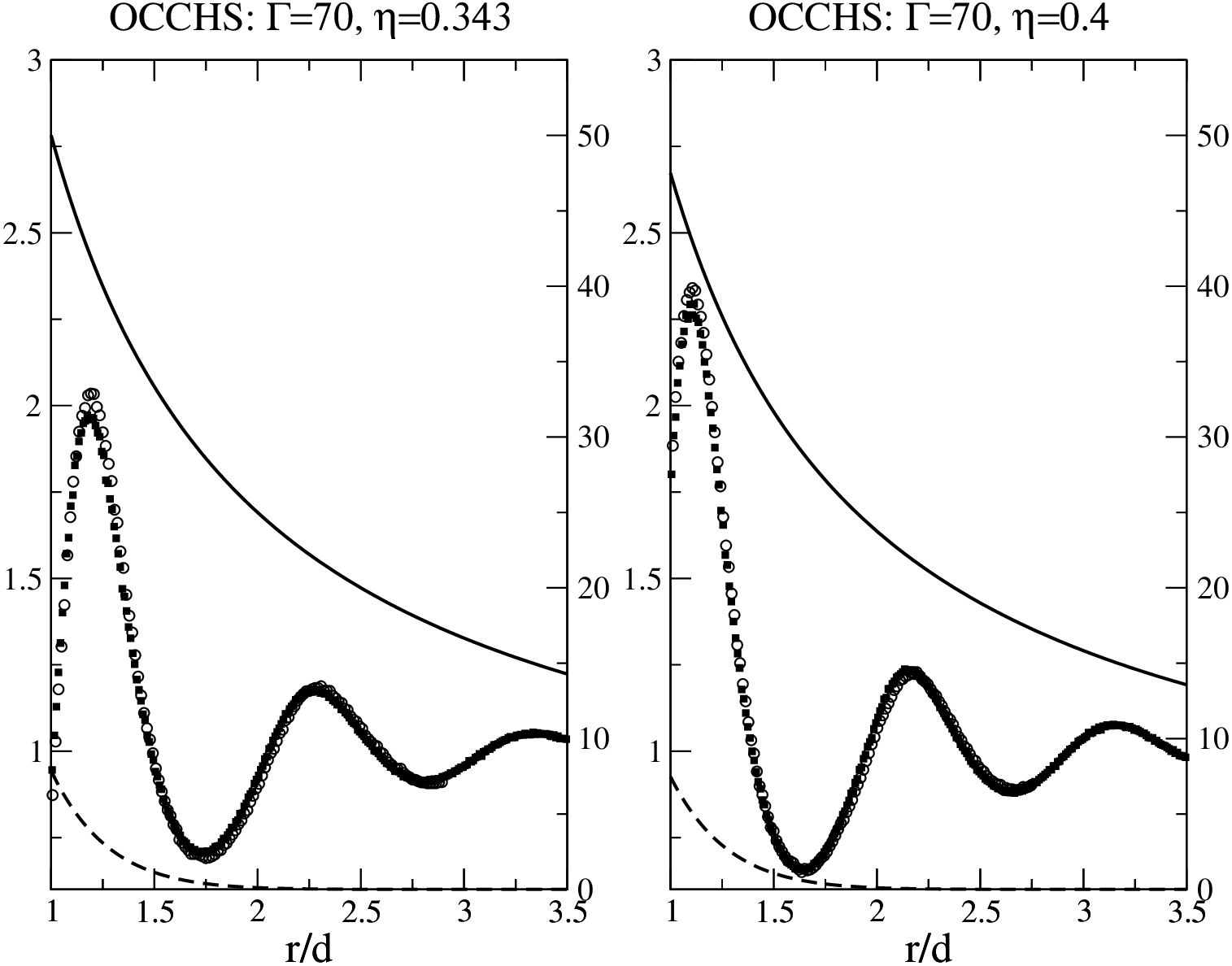}

\caption{\label{cap:occhsg20.gr} OCCHS correlation functions $g(r)$ at stronger
coupling strengths $\Gamma $ and large packing fractions $\eta $ for the full
and mimic systems as determined by MC simulations. Note the maximum in $g(r)$
away from contact for $\Gamma =70$, indicating the strength of the Coulomb
repulsions. Also shown are $\beta w(r),$ the full potential of the ionic
fluid, and the mimic potential $\beta u_{0}(r)$, both of which use the scale
on the right $y$-axis. There is a hard core interaction for $r<d$.}
\end{figure}

At high densities we expect that the local structure in $r$-space
of even the \emph{uniform} mimic system, where all corrections from
$\phi_{R1}$ are ignored, will closely resemble that of the full system
as suggested in eq \ref{eq:gog}. This is illustrated in Figure \ref{cap:occhsg20.gr}
for the OCCHS for high density states with moderately strong couplings.
Canonical Monte Carlo simulations are carried out to obtain the uniform
mimic system's correlation function $g_{0}(r)$,\cite{oursimulations} and these are compared
to previous simulations for the full system,\cite{hansen.jp.weis.jj:charged}
where Ewald sum methods
were used to account for the long ranged interactions.
Because of the short ranged interactions $u_{0}$ in the mimic system, our simulations
are completely straightforward and no Ewald sums or other special
treatment of the periodic boundary conditions are required.
Again a molecular-sized choice of $\sigma=d$ of order the nearest neighbor
spacing suffices.

Also shown on the same graphs are the bare ion potential $\beta w(r)$
and the mimic potential $\beta u_{0}(r)$. Despite the much smaller
amplitude of the latter and its much shorter range, the mimic $g_{0}(r)$
has a striking resemblance to the full system's $g(r),$ and the differences
are hardly visible on the scale of the graph. 

At the strongest couplings,
the correlation functions at both densities have a first peak shifted
away from contact with the embedded hard sphere. Such a correlation
function is very different from the correlation function of a hard
sphere fluid, which has its maximum at contact, and shows that the
strong short ranged parts of the Coulomb interactions can compete
with packing effects from hard cores even at high density. 
This also emphasizes the importance of having the softer piece
$w_{q0}(r)=q^{2}{\textrm{erfc}}(r/\sigma)/\epsilon r$
outside the hard core in our mimic system potential $\beta u_{0}(r)$
in eq \ref{eq:u0sigma} in order to reproduce correlation functions
in the OCCHS, especially for strong coupling states. 

There have been several
previous empirical attempts to fit correlation functions for Coulomb
systems at high density using effective short ranged systems. The DH
limit might suggest that a generalized Yukawa fluid could be useful,
\cite{leote-de-carvalho.rjf.evans.r:screened}
but the results were not very accurate and there was no systematic
way to determine parameters for the effective potentials.

Most relevant to our work are ion reaction field (RF) methods, where an effective
finite-ranged interaction $w_{q0}^{RF}(r)$ was originally determined from
the electrostatic potential of a positive point charge surrounded by a
neutralizing uniform spherical charge distribution with a radius
$r_c$. \cite{hummer1} Good results for correlation functions
were found using the RF method in several
applications at high density, though some spurious oscillations were
seen near the cutoff $r_c$. These were attributed to the discontinuity
of the second and higher order derivatives of $w_{q0}^{RF}(r)$ at $r_c$ and
better results were found using a smoother ``charge cloud'' distribution
that had discontinuities only in fourth and higher order derivatives at the
cutoff. \cite{hummer2}

Our $w_{q0}$ in eq \ref{eq:u0asymp} can be similarly interpreted
in terms of the potential arising from a positive point charge surrounded by a
neutralizing Gaussian charge distribution. All derivatives of $w_{q0}$ are
continuous because of the smooth cutoff, and by construction the associated
perturbation $w_{q1}$ decays very rapidly in $k$-space. It is the latter
property that fundamentally leads to mimic system behavior with a proper choice of
$\sigma$. Our work thus provides a conceptual framework for understanding
why RF methods can work as well as they do in some cases and how results can
be significantly improved, especially at lower densities or in nonuniform
environments by using the LMF theory.

\begin{figure}
\includegraphics[%
  scale=0.32]{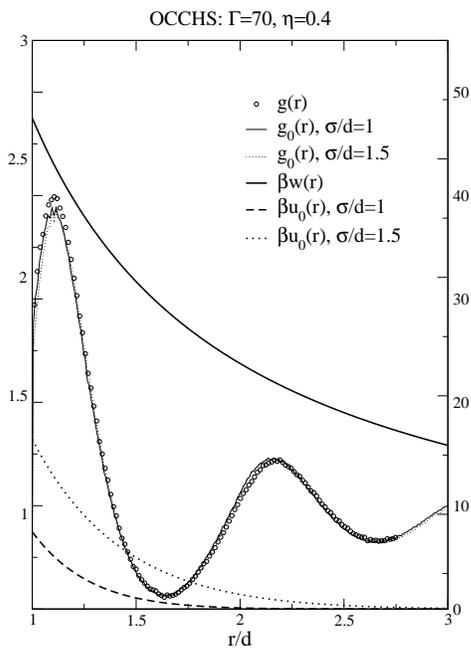}

\caption{\label{cap:occhsmoresig} Illustration of mimic system behavior. OCCHS
correlation function for the state $\Gamma =70$ and $\eta =0.4$ compared to those
of two different mimic systems with different $\sigma$ values.
Also shown are $\beta w(r)$, the full potential of the ionic
fluid, and the mimic potentials $\beta u_{0}(r)$, all of which use the scale
on the right $y$-axis.}
\end{figure}

Figure \ref{cap:occhsmoresig} gives a more detailed comparison of the structure
of the high density/strong coupling state with $\Gamma=70$ and $\eta=0.4$ to
that of two different mimic systems with $\sigma=d$ and $\sigma=1.5d$. Despite the
fact that the (repulsive) potential of the latter is always greater than or
equal to that of the former, both mimic systems have very similar correlation
functions that agree very well with that of the full system, which can be viewed
as the limit $\sigma=\infty$. Thus for high density states all mimic systems with
$\sigma>\sigma_{\min}$ have essentially the same short-ranged
structure in $r$-space.

\begin{figure}
\includegraphics[%
  scale=0.32]{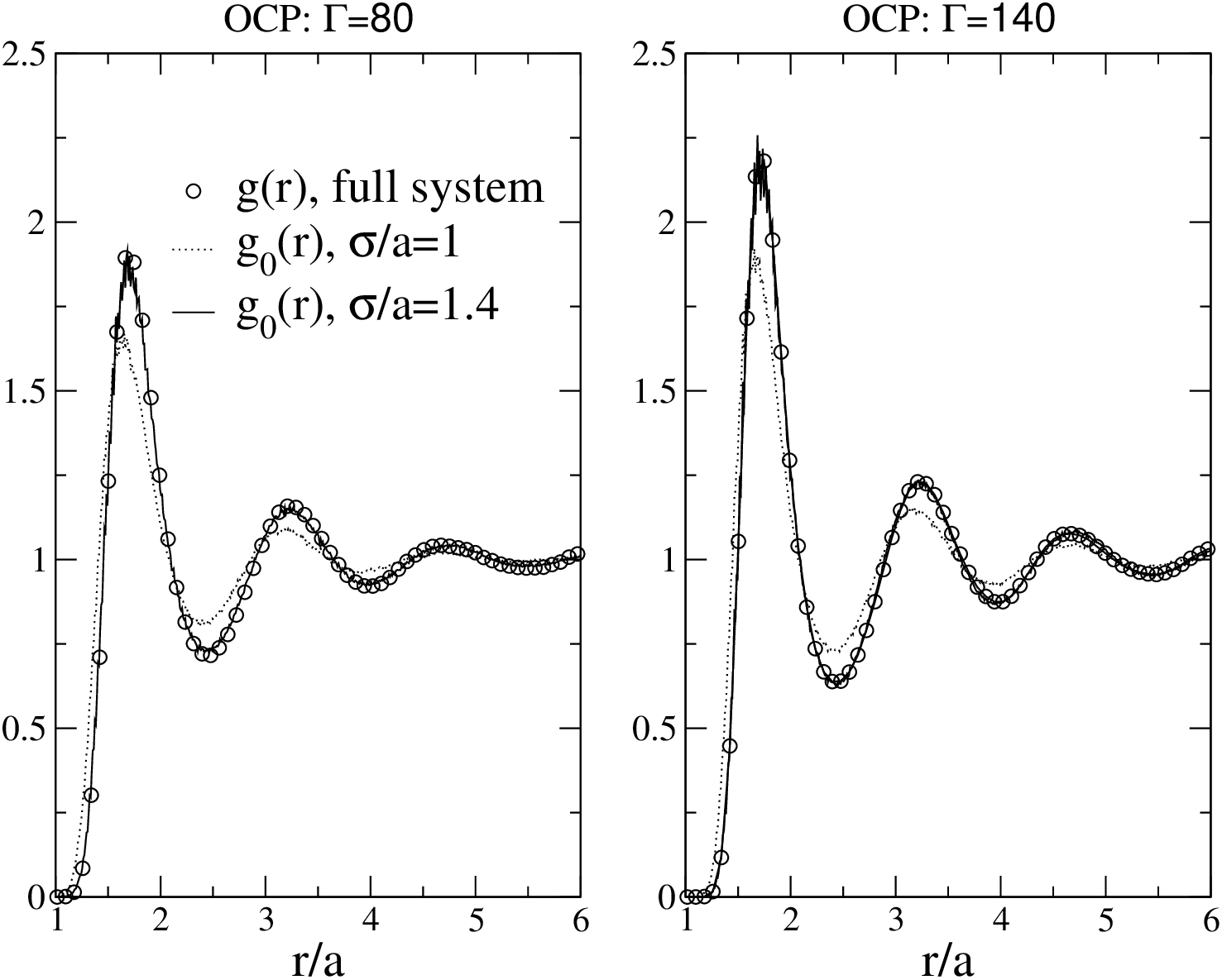}

\caption{\label{cap:ocphighg} High coupling strength OCP correlations for the full and
mimic systems. For $\sigma =1.4a$, $g_{0}(r)$ is essentially
indistinguishable from the full system's $g(r)$. However a smaller $\sigma =a$
fails to mimic the full system's correlations. Note that around $\Gamma
\simeq 170$ the OCP starts to freeze. \cite
{slattery.wl.doolen.gd.ea:n-dependence} Our simulations indicate that the
mimic system with $\sigma =1.4a$ also freezes at around the same $\Gamma $.}
\end{figure}

Figure \ref{cap:ocphighg}%
gives similar results for the OCP at very strong coupling strengths
$\Gamma=80$ and $\Gamma=140.$ We find excellent agreement with simulations
of the full system\cite{slattery.wl.doolen.gd.ea:n-dependence}
using a mimic system with $\sigma/a=1.4.$ As we
would expect, for small enough $\sigma$ the good agreement fails,
as illustrated by results for $\sigma/a=1.$

\begin{figure}
\includegraphics[%
  scale=0.32]{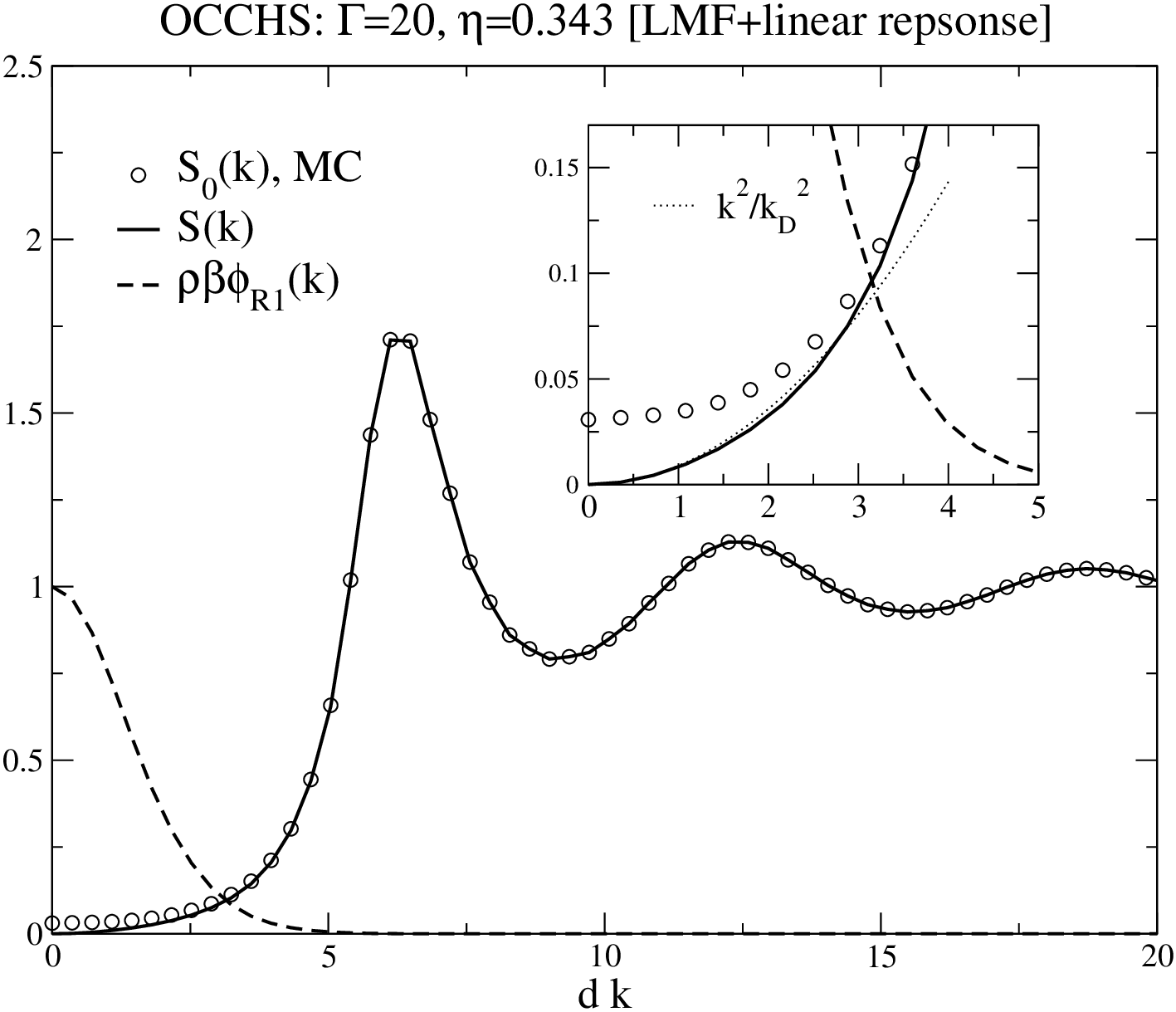}

\caption{\label{cap:occhsg20.sk} Linear response theory is used to approximate the
change in $S_R(k)=S(k)$ in the mimic system induced by the the perturbation
$\hat{\phi}_{R1}(k)$. With $\alpha =1,$ the linear response theory for $S(k)$ satisfies
both SL moment conditions. The inset is a blown-up view of the structure
factor at small $k$ where the differences in $S_{0}(k)$ and $S(k)$ can be
seen.}
\end{figure}

Figure \ref{cap:occhsg20.sk}
compares the mimic structure factor $S_{0}(k)$ and a simple estimate
for $S_{R}(k)=S(k)$ based on a linear response treatment\cite{linearresp}
of the effects of $\beta\rho^{B}\hat{\phi}_{R1}(k).$ Only at very small $k$ as
revealed in the inset can any differences be seen. The linear response
treatment turns out give an $S_{R}(k)$ that satisfies exactly both
the zeroth and second moment conditions with $\alpha=1$, and the
results converge to $S_{0}(k)$ at larger $k$ controlled by the factor
$\exp[-\frac{1}{4}(k\sigma)^{2}]$ arising from our choice of a mimic
system. These features would be found in any more exact treatment and
suffice at high densities to give a very accurate estimate of $S_{R}(k).$
This also suggests that the results from other approximate theories
may be improved by use of a good mimic system. Indeed, as already
shown in Figure \ref{cap:occhsg20.gr}, the simplest possible
theory where the effects from $\phi_{R1}$ are ignored completely,
already gives excellent results for short ranged correlations in
$r$-space. 

\subsection{Internal Energy at High Density}

\label{sec:energy}

With accurate approximations for $S(k)$ and $h(r)$ in hand, it is
straightforward to calculate thermodynamic properties by integration.
The simplest of these is the internal energy, given by eq \ref{eq:energy}.
One can always solve the LMF equation to obtain $h(r)$ and use this integral
to compute the internal energy of an ionic system, and we would expect
very accurate results. Here we show that because of the great similarity
of the local structure of the mimic and full systems in $r$- and
$k$-space at \emph{high density}, we can obtain an accurate estimate
of the energy in terms of the mimic system's energy and a simple analytic
correction without explicitly solving the LMF equation. 

Equation \ref{eq:energy} can be rewritten in $k$-space and the Coulomb
interaction separated into mimic and perturbation parts, so that
\begin{eqnarray}
\frac{\beta E^{ex}}{N} & = & \frac{1}{2}\frac{1}{(2\pi)^{3}}
\int d{\textbf{k}}\frac{4\pi\beta q^{2}}{\epsilon k^{2}}
[1-e^{-\frac{1}{4}(k\sigma)^{2}}]\rho^B\hat{h}(k)\nonumber \\
 &  & +\frac{1}{2}\frac{1}{(2\pi)^{3}}\int d{\textbf{k}}\frac{4\pi\beta q^{2}}{\epsilon k^{2}}
e^{-\frac{1}{4}(k\sigma)^{2}}\rho^B\hat{h}(k).\label{eq:split1}
\end{eqnarray}
In the first term of eq \ref{eq:split1}, because $\rho^B\hat{h}_{0}(k)$
differs from $\rho^B\hat{h}(k)$ only at small $k$, where the factor
$1-e^{-\frac{1}{4}(k\sigma)^{2}}$also approaches zero, we can replace
the latter by the former with little error. In the second term,
$e^{-\frac{1}{4}(k\sigma)^{2}}$ decays
very quickly at higher $k$, so only the small-$k$ features of $\rho^B\hat{h}(k)$
$\simeq-1+k^{2}/k_{D}^{2}$ from the SL moment conditions are needed
to have an accurate estimate of the integration.

The first term then
gives the internal energy of the mimic system
\begin{equation}
\frac{\beta E_{0}^{ex}}{N}=\frac{1}{2}\int d{\textbf{r}}u_{0}(r)\rho^B h_{0}(r),
\label{eq:E0}
\end{equation}
(with a background contribution $-\pi\beta q^{2}\sigma^{2}\rho^B/\epsilon$
arising from the use of $h_{0}$ in the integral rather than $g_{0}$),
while the second term corrects the mimic system's energy and can be
integrated analytically to give
\begin{equation}
\frac{\beta E_{1}^{ex}}{N}=\frac{\beta q^{2}}{\epsilon\sqrt{\pi}\sigma}
\left(-1+\frac{2}{(k_{D}\sigma)^{2}}\right).\label{eq:E1}
\end{equation}

Thus the internal energy can be estimated as
\begin{equation}
\frac{\beta E^{ex}}{N}\simeq\frac{\beta E_{0}^{ex}}{N}
+\frac{\beta E_{1}^{ex}}{N}.\label{eq:Eest}
\end{equation}
Results for this approximation are compared to MC results in Table
\ref{cap:occhsE.high}%
\begin{table}
\begin{tabular}{ccccccc}
\hline 
$\Gamma$&
$\eta$&
$\sigma /a$&
$\sigma /d$&
Full MC&
Theory [eq \ref{eq:Eest}]&
$\beta E_{0}^{ex}/{N}$\tabularnewline
\hline
20&
-&
1.4&
-&
-16.67&
-16.59&
-8.66\tabularnewline
20&
0.343&
1.4&
1&
-17.17&
-17.12&
-9.20\tabularnewline
20&
0.4&
1.4736&
1&
-17.33&
-17.27&
-9.73\tabularnewline
70&
-&
1.4&
-&
-60.81&
-60.72&
-32.65\tabularnewline
70&
0.343&
1.4&
1&
-61.09&
-61.02&
-32.94\tabularnewline
70&
0.4&
1.4736&
1&
-61.32&
-61.24&
-34.56\tabularnewline
80&
-&
1.4&
-&
-69.69&
-69.64&
-37.54\tabularnewline
125&
-&
1.4&
-&
-109.73&
-109.74&
-59.50\tabularnewline
140&
-&
1.4&
-&
-123.09&
-123.13&
-66.85\tabularnewline
160&
-&
1.4&
-&
-141.72{*}&
-141.57{*}&
-77.23\tabularnewline
\hline
\end{tabular}

\caption{\label{cap:occhsE.high}Excess internal energy $\beta E^{ex}/N$.
The full MC data are taken from references \onlinecite{hansen.jp.weis.jj:charged}
and \onlinecite{slattery.wl.doolen.gd.ea:n-dependence}.
At $\Gamma=160$, both the full and the mimic systems are near
solidification, and the results depend on initial conditions.}
\end{table}
for a variety of high density/strong coupling states. 
The Gaussian charge distribution is the key to the accuracy of eq
\ref{eq:Eest}, both because of its fast decay in $k$-space and
its use in revealing an excellent mimic system for the full ionic
system.

Note that at high densities or ionic strengths,
the Debye wave vector $k_{D}$ in eq \ref{eq:kD}
can be very large, making neutrality the major contribution to
$\beta E_{1}^{ex}/N\simeq-\beta q^{2}/(\epsilon\sqrt{\pi}\sigma)$.
Though the physical interpretation of this term is very different,
its limiting value is the same as the self-interaction correction
in the Ewald sum method. \cite{frenkelsmit}
A similar correction was used in ion RF methods. \cite{hummer1}

At high density where the compressibility $\chi_T$ of the mimic system
is very small, we can obtain a good estimate of the internal energy
by simply replacing $h(r)$ by $h_0(r)$ in eq \ref{eq:energy}, where because
of the background subtraction, a finite result is found for any short ranged
$h_0(r)$. Separating the Coulomb interaction as in eq \ref{eq:split1},
and using only the $k=0$ result $\rho^B\hat{h}_0(0)=-1+k_BT\rho\chi_T$ in
the second integral we find analogous to eq \ref{eq:E1}:
\begin{equation}
\frac{\beta E_{1}^{ex}}{N}\simeq\frac{\beta q^{2}}{\epsilon\sqrt{\pi}\sigma}
\left(-1+k_BT\rho\chi_T\right),\label{eq:E10}
\end{equation}
which agrees with eq \ref{eq:E1} in the relevant limit of
large $k_{D}$ and small $\chi_T$.
\section{Final Remarks}

\label{sec:conclusion}

It is straightforward to apply these idea to charge and size asymmetric
primitive models\cite{chenweeks} or to
``simple molten salt'' models\cite{hansenmac} with softer
repulsive cores. We have also used the LMF approximation to look at
the OCP near a charged hard wall.\cite{kaurchenweeks} Here very accurate results are found
both in the weak and strong coupling limits, and the more complicated
pair level theory recently introduced\cite{burak} is not required. Details will
be presented elsewhere, and we only make a few general remarks here. 

We have found for size asymmetric primitive models that the simplest
choice of a single $\sigma$ parameter for all species gives excellent
results. This can be understood as a consequence of Stillinger and
Lovett's fundamental insight that for general ionic mixtures universal
consequences of the long ranged Coulomb interactions can be seen in
the small wavevector behavior of the charge-charge correlation function.
By using the smeared charge distributions implied by a proper choice
of $\sigma$, we arrive at a smeared charge-charge correlation function
that has significant structure only at small wave vectors. The LMF
theory can then reproduce the exact long wavelength behavior found
by SL, and the slowly varying smeared Coulomb perturbations have little
effect on the shorter wavelength correlations induced by the modified
ion cores in the mimic system, just as illustrated here for the OCCHS. 

The main complication that arises for such mixtures is that the resulting
mimic systems will have short ranged attractive as well as repulsive
interactions, but this is required if we want the mimic system's
structure to resemble that of the full Coulomb system.
However, the LMF theory can then to a very good approximation take
care of the ``universal'' long ranged parts of the Coulomb interactions,
which cause major conceptual and computational problems in most approaches,
while leaving a nonuniversal, but surprisingly short ranged problem
to be treated by whatever means are available. Simulations are straightforward
and some recent theoretical developments for treating systems with
strong but short ranged interactions look very promising.\cite{prattquasi}
We believe the
ideas presented here offer a powerful general perspective, and
are actively pursuing their consequences for static and dynamic properties of
fluids with both short and long ranged interactions. 

\section{Acknowledgments}

We are grateful to Michael Fisher, Kirill Katsov, Michael Klein, and Lawrence Pratt for
helpful comments. We particularly want to thank J. J. Weis for 
sending us his simulation data for the strong coupling OCCHS correlation functions.
This work was supported by the National Science Foundation through Grant CHE-0111104. 

\appendix*
\section{Numerically stable version of LMF equation}

The LMF equation \ref{eq:lmfalpha} will produce the desired short ranged
$\phi_{R1}(r)$ with a finite value of $\hat{\phi}_{R1}(0)$ only
if $S_{R}(k)$ rigorously vanishes at $k=0$ as in eq \ref{eq:SR(k)expand},
which of course is the exact result. However such self consistent
equations are usually solved by iteration and any small errors in
an intermediate approximation to $S_{R}(k)$ at small $k$ are greatly
amplified. This can lead to numerical instabilities. The following
rewriting of eq \ref{eq:lmfalpha} can alleviate this problem.
We can remove the sensitivity at small $k$ by multiplying both sides
of eq \ref{eq:lmfalpha} by $k^{2}$, giving
\begin{equation}
k^{2}\beta\rho^{B}\hat{\phi}_{R1}(k)=\alpha k_{D}^{2}
\exp[-{\textstyle {\frac{1}{4}}}(k\sigma)^{2}]S_{R}(k).\label{eq:k^2phi}
\end{equation}
This equation suffices to determine $\hat{\phi}_{R1}(k)$ everywhere
except near $k=0,$ where $\hat{\phi}_{R1}$ is assumed to be regular.
We next formally write an identity involving $\hat{\phi}_{R1}(k)$
that remains finite as $k \rightarrow0$:
\begin{equation}
K^{2}\beta\rho^{B}\hat{\phi}_{R1}(k)
=K^{2}\beta\rho^{B}\hat{\phi}_{R1}(k),\label{eq:K^2phi}
\end{equation}
where $K$ is a (real) \emph{constant} wavevector. (More generally,
we can multiply both sides by a known real function of $k$ that does
not vanish as $k\rightarrow0.)$ We now add these two equations as written and
divide by $k^{2}+K^{2}$, which yields our final result \begin{eqnarray}
\beta\rho^{B}\hat{\phi}_{R1}(k) & = & \frac{\alpha k_{D}^{2}}{k^{2}+K^{2}}\exp[-{\textstyle {\frac{1}{4}}}(k\sigma)^{2}]S_{R}(k)\nonumber \\
 &  & +\frac{K^{2}}{k^{2}+K^{2}}\beta\rho^{B}\hat{\phi}_{R1}(k).\label{eq:lmfK}\end{eqnarray}
This equation has no divergences at small $k$ and is stable when
iterated, with a choice of $K$ of order $k_{D}$. A converged solution
will produce a $S_{R}(k)$ that vanishes identically at $k=0$ with
$\beta\rho^{B}\hat{\phi}_{R1}(0)$ finite. From eq \ref{eq:alpha2nd}
this quantity is in fact of order unity when the second moment condition
is satisfied. This way of rewriting equations of this kind
was suggested to us by Kirill Katsov.


\begin{thebibliography}{10}
\bibitem{potentialnorm}  This criterion can equivalently be reexpressed in
terms of the normalization of the electrostatic potential. \cite{outhwaite.cw:comment}
\bibitem{sigmachoice} Using the Debye-H\"{u}ckel approximation for the OCP
$S_{R}^{DH}(k)=k^{2}/(k^{2}+k_{D}^{2})$ and eq \ref{eq:kD}
in eq \ref{eq:lmfalpha}
(consistent with $\alpha =1$) we have
$\beta \rho ^{B}\hat{\phi}_{R1}^{D}(k)=3\Gamma e^{-\frac{1}{%
4}(k\sigma )^{2}}/[(ka)^{2}+3\Gamma ]$. The exact $S_{R}(k)$ has the
same form at small $k$ (in accord with the SL moment conditions) and
generally very different behavior at wavevectors larger than $k_{a}\equiv 1/a$
that characterize local structure in $r$-space. In order to achieve
mimic system behavior with $S_{R}(k)\simeq S_{0}(k)$ at larger wavevectors
we require $\beta \rho ^{B}\hat{\phi}_{R1}(k) \ll 1$ for all such wavevectors.
Since $S_{R}(k)$ is
generally of order unity, it should suffice to choose $\sigma $ so that
$\beta \rho ^{B}\hat{\phi}_{R1}^{D}(k_{a})=3\Gamma e^{-\frac{1}{4}(\sigma
/a)^{2}}/[1+3\Gamma ]\ll 1$. For large $\Gamma $ we require $\sigma $
of order $a$, but for small $\Gamma \ll 1$ even $\sigma =0$
will still satisfy this rough criterion.
\bibitem{oursimulations} The ${S}_0(k)$ at larger wavevectors can be accurately
determined by a Fourier transform of the simulated $g_0(r)$, but at small
wavevectors we directly sampled the ensemble average
\[
\frac{1}{N}<\hat{\rho}_{\mathbf{k}}\hat{\rho}_{-\mathbf{k}}>=
\frac{1}{N}<\mathop{\sum\sum}_{i\neq j}^{N}e^{-i\mathbf{k}\cdot(\mathbf{r}_{i}-\mathbf{r}_{j})}>+1.\]
We used $N=250$
and $432$ particles for the OCP and up to $686$ particles for the high
density OCCHS.
\end{thebibliography}
\end{document}